\documentstyle[11pt,newpasp,twoside,epsf]{article}
\def\edcomment#1{\iffalse\marginpar{\raggedright\sl#1\/}\else\relax\fi}
\reversemarginpar

\newcommand{\msun}{\mbox{M}_\odot}

\begin{document}
\title{The Lower Mass Function of Young Open Clusters}
 \author{J\'er\^ome Bouvier, Estelle Moraux}
\affil{Laboratoire d'Astrophysique, Observatoire de Grenoble, B.P.53, 38041
 Grenoble Cedex 9, France}
\author{John R. Stauffer}
\affil{SIRTF Science Center, California Institute of Technology, Pasadena,
 CA~91125,  USA}
\author{David Barrado y Navascu\'es}
\affil{LAEFF-INTA, Apdo. 50727, Madrid, Spain}
\author{Jean-Charles Cuillandre}
\affil{Canada-France-Hawaii Telescope Corp., Kamuela, HI 96743,  USA}

\begin{abstract}
  We report new estimates for the lower mass function of 5 young open
  clusters spanning an age range from 80 to 150 Myr.  In all studied
  clusters, the mass function across the stellar/substellar boundary
  ($\sim$ 0.072 $\msun$) and up to 0.4 $\msun$ is consistent with a
  power-law with an exponent $\alpha \simeq -0.5 \pm 0.1$, i.e., $dN/dM
  \propto M^{-0.5}$. 
\end{abstract}

\section{Introduction}

Young open clusters are ideal locations to search for isolated brown
dwarfs. Their youth ensures that substellar objects have not yet cooled
down to undetectable levels, and the rich stellar populations of the
nearest open clusters complement the recent discoveries of cluster brown
dwarfs to yield a complete mass function for coeval systems from the
substellar domain up to massive stars.

Nearby clusters have been surveyed by various groups in an effort to build
statistically significant samples of young brown dwarfs and derive reliable
estimates of the substellar mass function. In this contribution, we present
the latest results obtained from the CFHT Pleiades wide-field survey
(Section 2) and estimates of the lower mass function for several other open
clusters (Section 3). We then briefly discuss the potential effects of
cluster dynamical evolution on the shape of the mass function (Section 4).

\section{The CFHT 2000 Pleiades survey}

A Pleiades survey performed in 1996 with the UH 8K camera led to the a
preliminary estimate of the cluster mass function which was found to be
consistent with a power-law $dN/dM \propto M^{-0.6}$ over the mass range
from 0.04 to 0.3 $\msun$ (Bouvier et al. 1998). Proper motion measurements
and follow-up observations in the near-IR of the brown dwarf candidates
detected in this survey subsequently confirmed this estimate (Mart\'{\i}n
et al. 2000, Moraux et al. 2001).

\begin{figure}
\plottwo{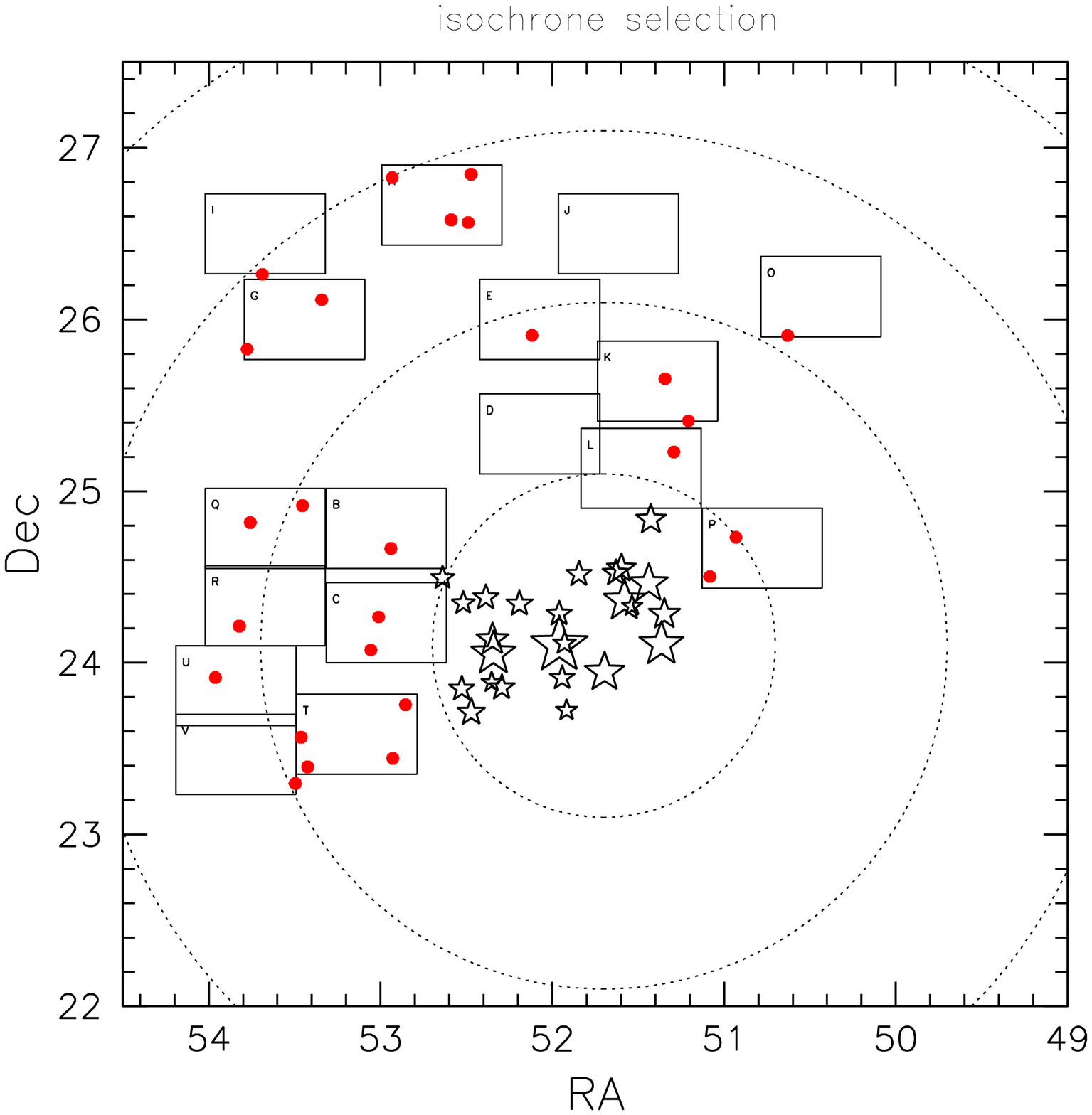}{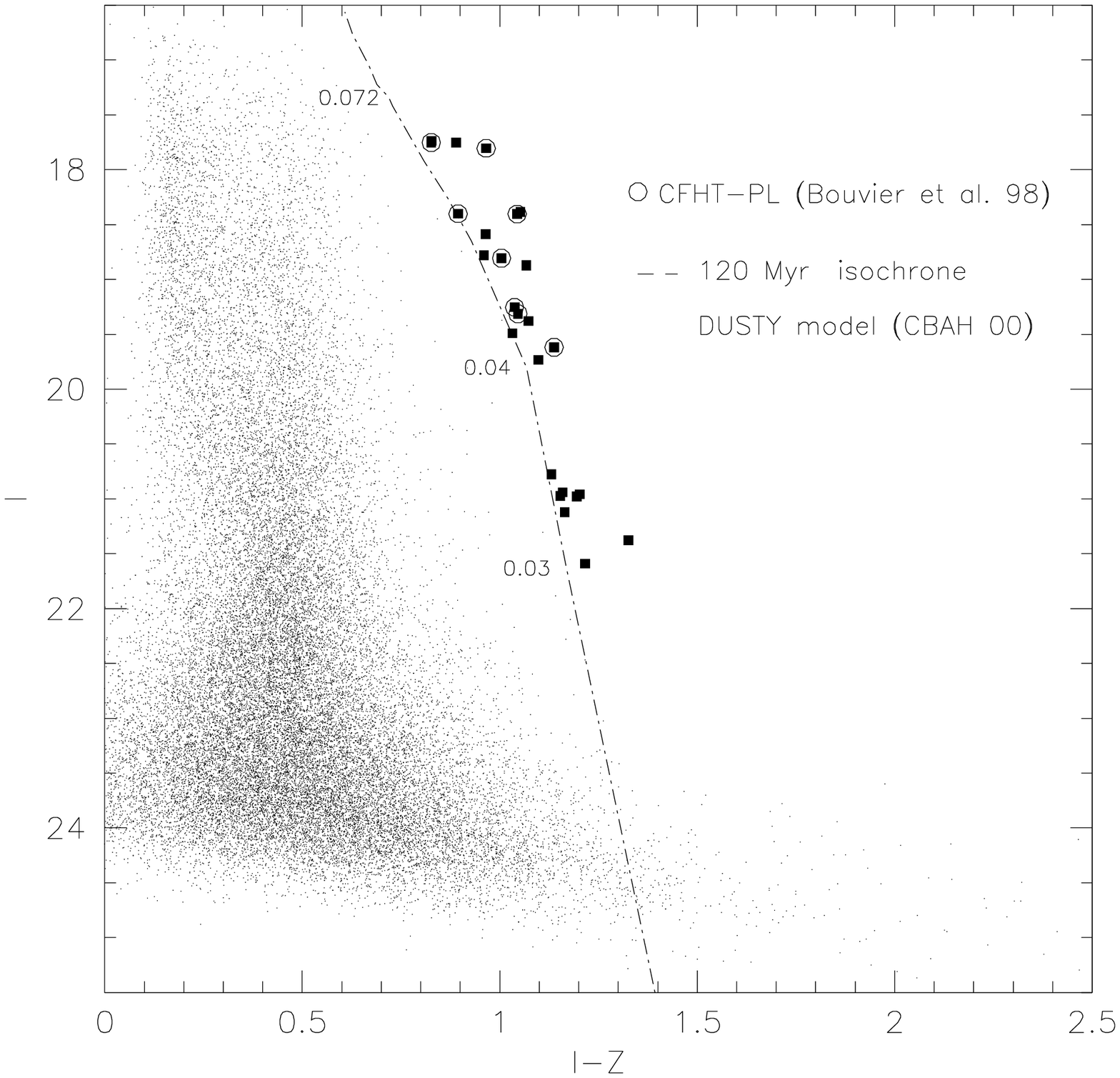}
\caption{ {\it Left Panel:} Area of the sky covered by the CFHT 2000 Pleiades
  survey. Each rectangle corresponds to one CFHT12K field. The star symbols
  indicate the 25 brightest stars of the cluster which have been avoided to
  prevent CCD saturation. Filled dots show the location of detected
  Pleiades brown dwarf candidates.  {\it Right panel:} A (I, I-z) CMD for
  all stellar-like objects identified on the long exposure images. The 120
  Myr isochrone of Chabrier et al.'s (2000) models shifted to the distance
  of the Pleiades is shown as a dash-dotted line labelled with mass (in
  $M_\odot$ unit).  Candidate brown dwarfs are identified as lying on or
  above the isochrone.  Brown dwarfs detected in the first CFHT survey and
  recovered here are shown as encircled squares while other candidates are
  new. (adapted from Moraux et al. 2002)}
\end{figure}

%

A new Pleiades survey was performed with the CFHT12K camera (Cuillandre et
al. 2000) in December 2000 covering a wider area (6.4 sq.deg.) and going
deeper (down to 0.025 $\msun$) than the original survey. Figure~1
illustrates the surveyed area on the sky (left) and the resulting (I, I-z)
color-magnitude diagram (CMD, right). New brown dwarf candidates are
selected from the CMD as lying on or above the 120~Myr isochrone of
Chabrier et al.'s (2000) models.  Pending follow-up observations of the
lowest-mass candidates, the photometric selection of substellar objects is
stopped at 0.030 $\msun$ where contamination by field stars becomes
significant.

\begin{figure}
\plotone{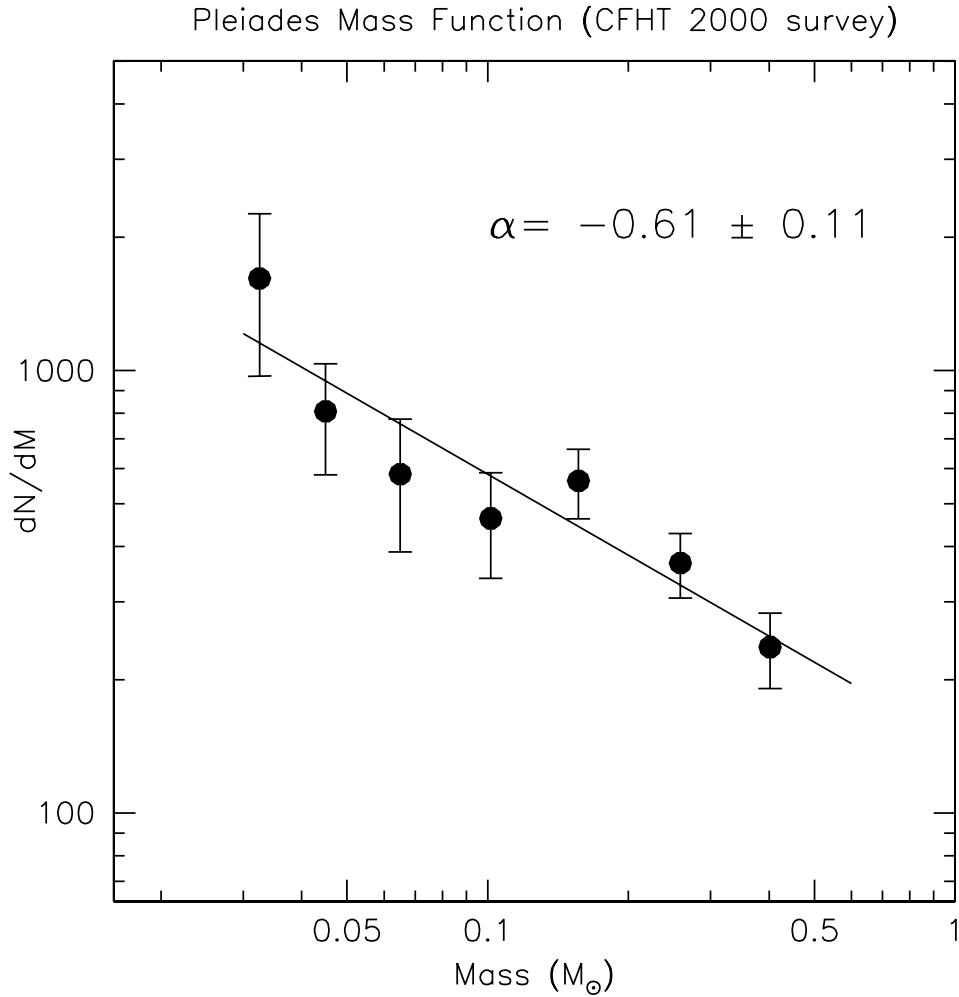}
\caption{The Pleiades mass function across the stellar/substellar
  boundary. Note that all data points are derived from the same survey,
  using short exposures for the stellar domain and long exposures for the
  substellar regime. This provides a consistent determination of the slope
  of the cluster's mass function in the mass range from 0.030 to 0.4
  $\msun$. A least-square fit to the data point yields a power-law exponent
  $\alpha = -0.61 \pm 0.11$. (adapted from Moraux et al.  2002)}
\end{figure}

Taking into account the contamination of photometrically-selected brown
dwarf candidates by field dwarfs (cf. Moraux et al. 2001), the number of
brown dwarfs detected in each magnitude bin is directly converted into a
mass function using the (I magnitude, Mass) relationship of Chabrier et
al.'s (2000) 120~Myr model. The mass function (MF) derived for the cluster
across the stellar/substellar boundary is shown in Figure~2. The data
points in the stellar domain have been derived from short exposures while
the derivation of the MF in the substellar domain is obtained from the long
exposures CMD shown in Figure~1. Over more than a decade in mass, from
0.030 to 0.4 $\msun$, the observed mass function is reasonably well-fitted
by a power-law $dN/dM \propto M^{-0.61 \pm 0.11}$, consistent with previous
determinations (Bouvier et al. 1998, Hambly et al. 1999, Mart\'{\i}n et al.
2000, Moraux et al. 2001).

\section{The lower mass function of Pleiades-age open clusters}

Apart from the Pleiades, a handful of other galactic open clusters are
young and close enough to allow deep brown dwarf searches (cf.  Stauffer \&
Barrado y Navascu\'es, this volume). Figure~3 shows current determinations
of the lower MF for M35 ($\sim$150 Myr, Barrado y Navascu\'es et al., in
prep.), Alpha Per ($\sim$80 Myr, Barrado y Navascu\'es et al. 2002) and for
NGC 2516 and Blanco 1 ($\sim$150 Myr, Moraux et al., in prep.).

\begin{figure}
\plotone{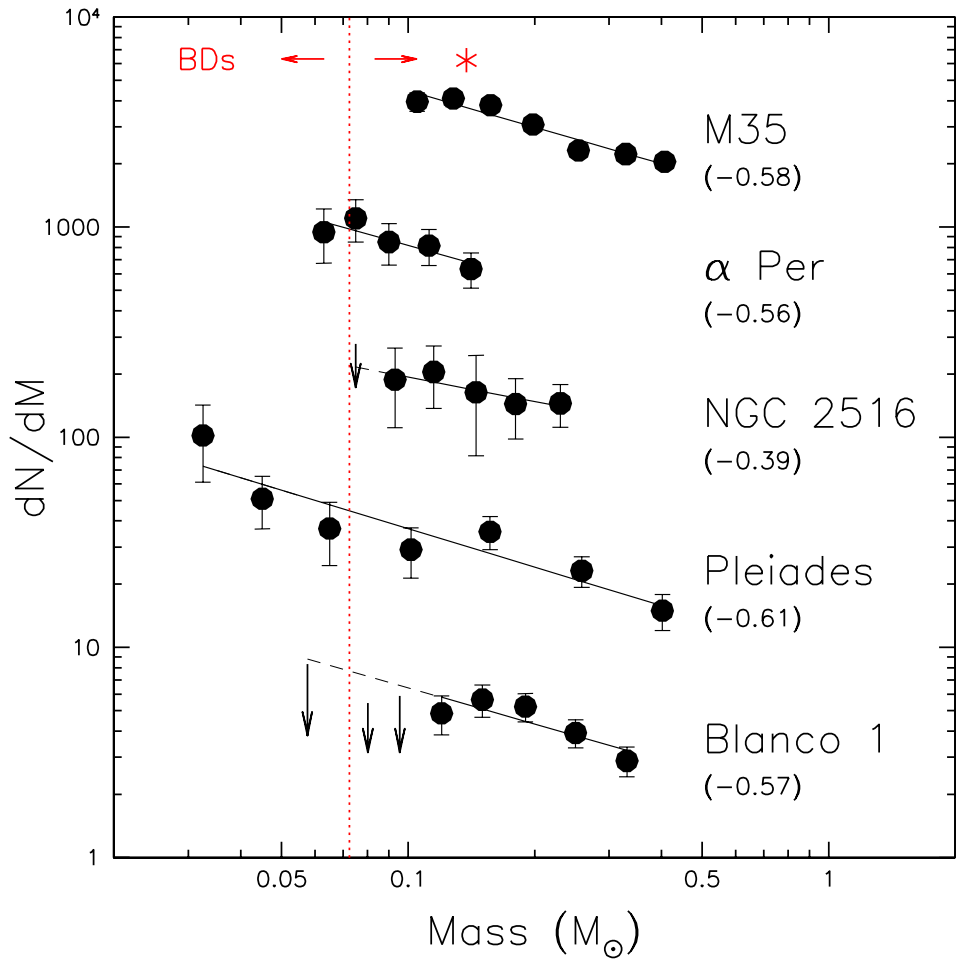}
\caption{The mass function of young open clusters across the
  stellar/substellar boundary. Each cluster is identified by its name and
  the index $\alpha$ of a power-law fit to the mass function ($dN/dM
  \propto M^\alpha$, shown as a solid line) is given within parantheses.
  Vertical arrows at the low mass end of NGC 2516's and Blanco 1's MF
  indicate 3$\sigma$ upper limits. dN/dM of each cluster has been
  vertically scaled for clarity.}
\end{figure}

The MF estimates have been derived from optical CMDs in a similar fashion
as for the Pleiades (see above) over a mass range extending from low mass
stars down to the stellar/substellar boundary and below it for the
youngest/nearest clusters. When approximated by a power-law, the slope of
the MF over this restricted mass range is strikingly similar for the
various clusters, with a power-law index $\alpha \simeq -0.5 \pm 0.1$ within
uncertainties.

The main uncertainties associated with these preliminary results are
related to small number statistics (shown as Poisson error bars in
Figure~3) and, for some clusters, to the contamination of the
photometrically selected BD candidates by field dwarfs. Follow-up
observations are being acquired to confirm the BD status of some of the
photometric candidates. In the meantime, vertical arrows drawn in Figure~3
at the low mass end of the MF indicate the current 3$\sigma$ upper limit to
the total number of candidates.

Pending the resolution of these uncertainties, there is presently no
evidence for significant differences in the mass function between the
Pleiades cluster itself and other Pleiades-type open clusters studied so
far across the stellar-substellar boundary.

\section{Dynamical effects and the origin of brown dwarfs} 

The main objective in determining the lower mass function is to constrain
the star and brown dwarf formation process(es). A pressing issue is then
whether the MF derived for young open clusters is representative of the
initial mass function (IMF), i.e. the distribution of the masses of
condensed objects resulting from their formation process.  In other words,
is the cluster population observed at an age of about 100 Myr
representative of the initial population of the cluster at the time it
formed~?

If the details of the star-formation process are not relevant to the
subsequent evolution, and if brown dwarfs are formed like stars, then the
difference between the IMF and the observed MF is only due to "classical"
dynamical evolution.  As the cluster evolves, dynamical processes act to
deplete its lowest mass members. Weak gravitational encounters lead to mass
segregation and to the evaporation of the lowest mass objects. Yet, current
models predict that only about 10\% of the low mass stars and brown dwarfs
will have escaped a Pleiades-like cluster at an age of 100 Myr by this
process (e.g., de la Fuente Marcos \& de la Fuente Marcos 2000).
Furthermore, the predicted loss rate is nearly the same for substellar
objects and low mass stars (since there is no dynamical boundary between
the stellar and substellar regimes), so that the shape of the mass function
will not be affected across the stellar/substellar boundary. Hence, the
secular dynamical evolution of stellar clusters is not expected to
significantly deplete the IMF at low masses during the early evolutionary
stages.

However, in some theoretical models, dynamical processes intimately linked
to the star-formation process are predicted to significantly influence the
observed MF at Pleiades age. In the scenario Reipurth \& Clarke (2001)
proposed for the formation of isolated brown dwarfs, the lowest mass
fragments of protostellar aggregates are dynamically ejected and may
rapidly leave the cluster if their ejection velocity exceeds the escape
velocity.  Several models of dynamical ejection have been developped along
these lines and all suggest ejection velocities of order of a few km/s,
though the output of each particular model differs in the details (cf.
Sterzik \& Durisen 1998, Bate et al.  2002, Delgado et al., in prep.).

\begin{figure}
\plotone{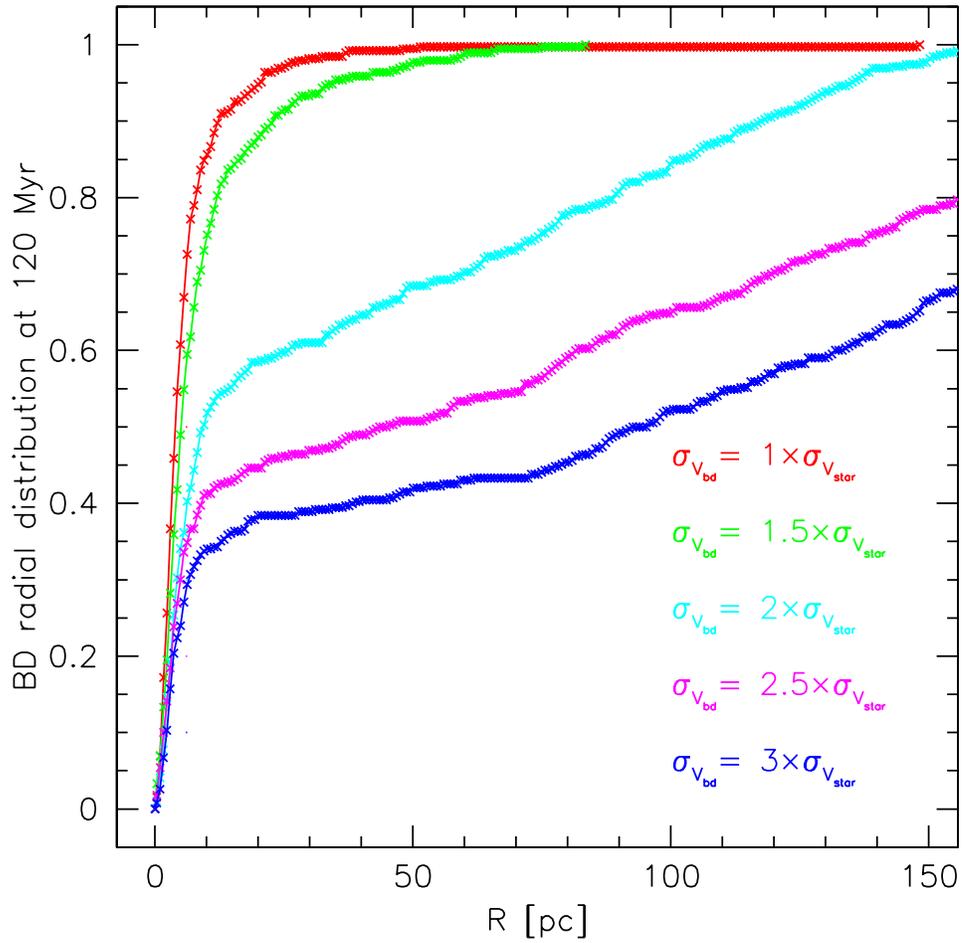}
\caption{The radial distribution of brown dwarfs at an age of 120 Myr
  computed from the numerical simulation of the dynamical evolution of a
  Pleiades-like cluster. From top to bottom the cumulative distributions
  illustrate the effect of increasing the initial velocity dispersion of
  cluster BDs. Note that the Pleiades cluster, with a tidal radius of about
  16 pc, is entirely contained in the first 2 bins on the x-axis.}
\end{figure}

Following this scenario, Figure~4 shows the results of a N-body 2 numerical
simulation of the dynamical evolution of a Pleiades-like cluster (Moraux \&
Clarke, in prep.). The simulation includes 1600 objects whose mass
distribution follows a prescribed mass function, with the initial velocity
dispersion of brown dwarfs being scaled relative to that of stars, and
assuming the whole system is virialized.  The various curves in Figure~4
illustrate the computed spatial distribution of brown dwarfs at an age of
120 Myr depending on their initial velocity distribution.  Note that the
Pleiades cluster (r$_{tidal} \simeq$ 16 pc) is entirely contained in the
very first 2 bins. It is seen that half or more of the initial cluster
brown dwarfs will have left the cluster by an age of 120 Myr if their
initial velocity dispersion exceeds that of stars by a factor of 2.

Such dynamical processes may be expected to significantly modify the lower
mass function of young open clusters in a way which sensitively depends
upon the specific properties of each cluster, such as initial stellar
density and radius, thus presumably leading to different lower MF for
different clusters. Yet, no significant difference is found so far between
the lower MF of the studied clusters. Also, the lower MF of young open
clusters appears similar to that of star forming regions such as Orion
Trapezium (Luhman et al. 2000) or $\sigma$ Ori (B\'ejar et al. 2001).
These results then suggest that dynamical processes are not in fact
predominant in the formation and early evolution of open clusters and that
the MF derived for the Pleiades and Pleiades-like clusters may indeed be
representative of the IMF.

\section{Conclusions} 

Deep wide-field photometric surveys of brown dwarfs in nearby young open
clusters have yielded estimates of the mass function across the
stellar/subtellar boundary. The best studied cluster so far is the Pleiades
whose lower mass function can be approximated by a power-law with an
exponent $\alpha = -0.6 \pm 0.1$ (i.e.  $dN/dM \propto M^{-0.6}$) over the
mass range 0.03-0.4 $\msun$. Though the determination of the mass function
in other Pleiades-age clusters (M35, Alpha Per, NGC 2516, Blanco 1) is not
yet as precise as for the Pleiades itself, current estimates suggest that
there is no appreciable differences in the shape of the lower MF between
the various clusters, regardless of their precise age, metallicity or
richness. This might be an indication that the currently measured mass
function of these clusters at an age of about 100 Myr is representative of
their initial mass function (IMF) and thus provides a quantitative
constraint to the formation scenarios for stars and brown dwarfs.

\end{document}